\documentclass[onecolumn,pdflatex,nofootinbib,superscriptaddress,floatfix]{revtex4-1}
\usepackage{graphicx}
\usepackage{bm}
\usepackage{placeins}
\usepackage{xspace}
\usepackage{bm,amsmath,amssymb,amsfonts,gensymb}
\usepackage{comment}
\usepackage{hyperref}
\hypersetup{
    colorlinks=true,
    linkcolor=blue,
    citecolor=blue,
    filecolor=blue,      
    urlcolor=blue,
}
\usepackage{cleveref}

\usepackage[russian,english]{babel}

\newcommand{\be}{\begin{equation}}
\newcommand{\ee}{\end{equation}}

\usepackage{color}

\usepackage[english]{babel}
\begin{document}

\title{Singularities in static spherically symmetric configurations of General Relativity with strongly nonlinear scalar fields
 }
\author{O.~S.~Stashko}
\affiliation{Taras Shevchenko National University of Kyiv, Ukraine}
\author{V.~I.~Zhdanov}
\affiliation{Taras Shevchenko National University of Kyiv, Ukraine}

\date{\today}

\pacs{11}


\begin{abstract}
There are a number of publications on relativistic objects  dealing either with black holes or naked singularities in the center. Here we show that there exist static spherically symmetric solutions of  Einstein equations with a strongly nonlinear scalar field with potential $V(\varphi)\sim\sinh(\varphi^{2n})$, which allow the appearance of singularities of a new type  (``spherical singularities'')  outside the center of isolated configuration. The space-time is assumed to be asymptotically flat. Depending on the configuration parameters, we  show  that the distribution of the stable circular orbits of test bodies around the configuration  is  either   similar to that in the case of the  Schwarzschild solution (thus mimicking an ordinary black hole), or it  contains additional rings of unstable orbits.
\end{abstract}
\maketitle
\section{Introduction}\label{introduction}
 Nowadays, the~concept of a black hole has become a common element of astrophysical research~\cite{1973blho.conf..343N,Antonucci1993,Bianchi_2012}. 
However, in~order to be fully confident in the theoretical idea, it must be compared with alternative models.  This is one of the motivations for many theoretical works devoted to non-canonical and even exotic solutions of General Relativity and its modifications, which could  have similar astrophysical consequences. The~non-exhaustive list below includes  papers on various aspects of compact objects with  naked \mbox{singularities~\cite{2014CQGra..31s5013S,Shao_2021,Chakraborty_2019,2020PhRvD.101d3005B,Pugliese_2011,Pugliese_2013,Joshi_2013,Shahidi_2020,2016PhRvD..93b4024B,2018Stashko, Bambhaniya_2019},} with wormholes~\cite{Karimov_2019,paul2019observational,narzilloev2021particle,Abdujabbarov_2009,2014PhRvD..90b4071L}, boson stars~\cite{2016Vincent,Grandclement_2014,LieblingL,Lamy_2018, Herdeiro_2021}, and the other non-singular objects mimicking the black holes~\cite{Dymnikova_2019,Stuchl_k_2015,Schnenobach2014}. 
The question  arises as to how the aforementioned exotic configurations were formed; probably, not all corresponding  models can reflect real astrophysical situations. However, final answers about the reality of these models must be based on observations. This explains why the interest in the ``black hole mimickers'' greatly increased after the Event Horizon Telescope observed the image of the accretion disk around the supermassive black hole in the center of M87~\cite{Akiyama2019}.  Correspondingly, much attention is paid to the observational manifestations of different models dealing with motion of test particles and photons in the gravitational field of the compact \mbox{object~\cite{Chowdhury_2012,2014CQGra..31s5013S,Stuchl_k_2015,Zhou_2015,Chakraborty_2019,Dymnikova_2019};} an  important point concerns the images of   accretion disks, radiation fluxes,  forms of the relativistic Fe $K\alpha$ line~\cite{Shaikh_2019,Gyulchev_2019,Gyulchev_2020,2021arXiv210614697G,Collodel2021,Sau_2020,Schnenobach2014,Cao_2016,2018PhRvD..98d4024Y,2017JCAP...08..014S,2018JCAP...08..044L}, etc.

A number of models are based on static solutions of the Einstein equations with scalar field (SF) \cite{Chowdhury_2012,Zhou_2015,Shaikh_2019,Gyulchev_2019,Gyulchev_2020,Mart_nez_2004,Sau_2020}. These papers use analytically defined metrics, in~particular the Fisher--Yanis--Newman--Winicour solution \cite{Fisher,JNW,Wyman,Virbhadra_1997}.
Papers~\cite{Sau_2020,Shaikh_2019,Gyulchev_2019,Gyulchev_2020,Chowdhury_2012,Zhou_2015} deal with singularities in the center, the~solutions used do not have other   ``physical'' singularities  outside the center. 
It was proved~\cite{ZhdSt} that this is  a fairly general situation, at~least when considering an isolated  spherically symmetric configuration  of General Relativity with a scalar field. Here the gravitational field suppresses the appearance of ``spherical  singularities'' and only a naked singularity in the very center is possible.  The~proof uses an assumption that the SF self-interaction potential is exponentially bounded~\cite{ZhdSt}. Thus, the~singularity for any non-zero value of the radial variable  in the curvature (Schwarzschild) coordinates  is  prohibited and  the question arises whether this result will be preserved  in the case of a sharper dependence of the SF potential for the  large field values? This is a key point of the present~paper. 

In this article, our goal is to provide examples showing that in the case of a sufficiently strong nonlinear behavior of the potential (with a sufficiently fast growth rate), spherical singularities (SS) can arise. To~relax the restrictions of~\cite{ZhdSt},  we choose $V(\phi)=\sinh(\phi^{2n})$, $n>2$ as a representative of such potentials. This is a technically convenient choice: for large $\phi$, this potential grows faster  than any exponentially bounded function; on the other hand, for~small $\phi$ it behaves as a monomial potential that allows us to use earlier  results concerning asymptotic behavior $\phi(r)\sim r^{-1}$ for $r\to\infty$.

The paper is organized as follows. In~Section~\ref{mathematical} we present initial relations. In~\mbox{Sections \ref{asymptotic_behavior} and \ref{numerical_solutions}} we show the  possibility of ``spherical singularities'' and present numerical solutions for the Einstein equations with SF in the case of spherical symmetry. In~Section~\ref{Test_particle_motion}  we  study  distributions of the stable circular orbits (SCO), which is important to study a thin  accretion disk around the configuration. We also  estimate the  radiation flux from the  accretion disk within the Page--Thorne  model~\cite{Page_Thorne}.

\section{Initial~Relations}\label{mathematical}
We consider a static spherically symmetric  space-time   metric  in the curvature coordinates, that is
\begin{equation}\label{metric}
    ds^{2} =e^{\alpha } dt^{2} -e^{\beta } dr^{2} -r^{2} \left[ d\theta^{2}+\sin^2\theta d\varphi^2 \right] ;
\end{equation}
  this fixes the radial variable $r>0$.

The gravitational field interacts with real SF 
 $\phi(r)$   described by  Lagrangian density
\begin{equation}\label{lagrangian}
L=\frac{1}{2} \partial_{\mu}\phi \partial^{\mu}\phi -V(\phi);
\end{equation}
 the SF potential   is
\begin{equation} \label{potential} 
V(\phi)=\sinh(\phi^{2n } ) .
\end{equation} 

This potential is strongly nonlinear  for $\phi\to\infty$ and for $n\ge 1$ it grows faster than $|\phi|^a \exp(b\phi)$ for any $a,b$. 

The  SF equation following from (\ref{lagrangian}) is
\begin{equation} \label{EE3} 
\frac{d}{dr} \left[r^{2} e^{\frac{\alpha -\beta }{2} } \frac{d\phi }{dr} \right]=r^{2} e^{\frac{\alpha +\beta }{2} } V'(\phi ) 
\end{equation} 

The independent from (\ref{EE3}) Einstein equations are reduced to the form (see, e.g.,~\cite{ZhdSt})
\begin{equation} \label{EE1} 
\alpha '+\beta '=8\pi r\phi '^{2}  ,             
\end{equation}
\begin{equation} \label{EE2} 
\beta '-\alpha '=\frac{2}{r} \left(1-e^{\beta } \right)+16\pi re^{\beta } V(\phi ),                     
\end{equation}

We focus on isolated systems with mass $M$ in the  asymptotically flat space-time assuming for $r\to\infty$
\begin{equation} \label{asy} 
\mathop{\lim }\limits_{r\to \infty } \left[r {\alpha(r) } \right]=-\mathop{\lim }\limits_{r\to \infty } \left[r\beta(r)\right]=-r_{g},~~ r_{g} =2M>0.       \end{equation} 

As for SF, we assume $\phi (\infty )=0$; then  for  (\ref{potential}) we have $V(\phi)\approx \phi^{2n}$ for $r\to\infty$ and therefore we can use some of the results on monomial potentials from~\cite{ZhdSt,stashko2021accretion}, concerning asymptotic behavior for large $r$:
\[\phi(r)\sim \exp(-\mu r)/r^{1+\mu M}\]
for $n=1$ (linear massive scalar field     with mass $\mu$ \cite{Stashko_Zhdanov_2019a});
\begin{equation}\label{asympt_inf}
\phi(r)\sim \left\{\frac{1}{r^{1/(n-1)}}, \,\, 1<n<2;\quad
\frac{1}{r \sqrt{|\ln r|}},\,\,n=2;\quad \frac{1}{r},\,\,n>2 ;\right\} 
\end{equation}
(see Appendix A of~\cite{ZhdSt}); asymptotic relations for $\phi'(r)$ can be obtained by formal differentiation of (\ref{asympt_inf}).

\section{ Asymptotic Behavior Near~Singularity}\label{asymptotic_behavior}
First of all we note that for regular  solutions of (\ref{EE3})--(\ref{EE2})  within some interval $(r_0,\infty),$ \mbox{$\,r_0>0$,}  satisfying  conditions (\ref{asympt_inf}) with non-trivial $\phi(r)$, we can show that functions $\phi(r)$ and $\phi'(r)$ preserve their signs. Indeed, for~potential (\ref{potential}) inequality $\phi V'(\phi)>0$ is valid for $\phi\ne 0$, and~using Equation~(\ref{EE3}) we get
\begin{equation}\label{negative_phi_phi'}
  \frac{d}{dr} \left[r^{2} e^{\frac{\alpha -\beta }{2} }\phi\, \frac{d\phi }{dr} \right]=r^{2} e^{\frac{\alpha +\beta }{2} }\phi\, V'(\phi ) + r^{2} e^{\frac{\alpha -\beta }{2} } \left[\frac{d\phi }{dr}\right]^2 >0.  
 \end{equation}
 
Therefore, function $r^{2} e^{\frac{\alpha -\beta }{2} }\phi\, \phi' $ is monotonically increasing. On~account of  conditions \mbox{(\ref{asy})} and (\ref{asympt_inf}), it is strictly negative  for large $r$; therefore it cannot be equal to zero  and so is   $\phi(r)\phi'(r)<0$. Whence we infer that functions $\phi(r),\,\phi'(r)$ do not change their~signs. 

Further, for~definiteness, we assume $\phi(r)>0$, $\phi'(r)<0$. 

Now we turn to the singularities at some $r_s>0$ in case of potential (\ref{potential}). We are looking for solutions on $(r_s,\infty)$ such that
\begin{equation}
\label{inf_1}
  \phi'(r)\to -\infty,~ r\to r_s +0,
\end{equation}
for some $r_s >0$. Our aim is to estimate asymptotic properties of these~solutions.

  Equations \eqref{EE1} and \eqref{EE2} yield
\begin{equation}\label{(1+2_sum)}
    \beta '=4\pi r\phi '^{2} +\frac{1}{r} \left(1-e^{\beta } \right)+8\pi re^{\beta } V(\phi ).
\end{equation}

Numerical simulations near singularity suggests that $\alpha (r)$ is a slowly varying function. So, as~a first approximation, we are neglecting this function compared to $ \beta (r) $.   Under~this assumption we get for the leading terms of  Equation \eqref{EE3}
\begin{equation}
    e^{-\beta /2} \frac{d}{dr} \left[e^{-\beta /2} \phi '\right]\simeq V'(\phi ).
\end{equation}

This is a  rough approximation that is valid in very small interval near the singularity. We have then $e^{-\beta } \phi '^{2} \simeq 2V(\phi )+const$, where the constant will be neglected in comparison with $V(\phi )$ for $r\to r_s +0$, so that
\begin{equation} \label{GrindEQ__4_} 
e^{-\beta } \phi '^{2} \simeq 2V(\phi ) .                                                    
\end{equation} 

In view of (\ref{inf_1}), $e^{\beta } V(\phi )\simeq \phi'^2\to \infty $ for $r\to r_s +0$ and then it is easy to see that the principal terms on the right hand sides of \eqref{EE1} and \eqref{EE2} are asymptotically the same. This justifies our  assumption about  $\alpha (r)$. 

Substitution into (\ref{(1+2_sum)}) yields
\begin{equation}\label{(4++)}
    \beta '\simeq 16\pi re^{\beta } V(\phi ),
\end{equation}
where we discarded the lower order terms. This allows us to reduce the problem to the system of two equations~.

Then $\beta (r)$ is monotonically increasing  (for $r\in (r_s ,r_{1} ]$, where $(r_{1} -r_s )/r_s\ll 1$).
For $r\to r_s +0$ Equation~(\ref{(4++)}) yields
\begin{equation} \label{GrindEQ__5_} 
\frac{d}{dr} \left[e^{-\beta /2} \right]\approx -8\pi r_s e^{\beta /2} V(\phi ).                                        
\end{equation} 
From Equation \eqref{GrindEQ__4_}  we have
\begin{equation} \label{GrindEQ__6_} 
\phi '\approx -\sqrt{2} e^{\beta /2} \sqrt{V(\phi )} ,                                             
\end{equation} 
where we take into account that $\phi (r)$ is~decreasing. 

 Dividing \eqref{GrindEQ__5_} by \eqref{GrindEQ__6_} we have equation 
\[\frac{d}{d\phi } \left[e^{-\beta /2} \right]=\frac{8\pi r_s }{\sqrt{2} } \sqrt{V(\phi )} \] 
that can be solved in quadratures. 
In the leading terms for  $r\to r_s +0$
\begin{equation} \label{exp-beta} 
e^{-\beta (r)/2} \simeq \frac{8\pi r_s }{\sqrt{2} } \Phi (\phi )+e^{-\beta (r_{1} )/2} \sim \frac{8\pi r_s }{\sqrt{2} } \Phi (\phi ) 
\end{equation} 
where we denote
\[\Phi (\phi )=\int _{\phi (r_{1} )}^{\phi }\sqrt{V(x)}  dx=\int _{\phi (r_{1} )}^{\phi }\exp \left(\frac{1}{2} x^{2n} \right) dx.\] 

This can be expressed by the incomplete gamma-function, leading to the asymptotic formula:
\begin{equation}  
\label{GrindEQ__8_}
    \Phi (\phi )=\frac{1}{n\phi ^{2n-1} } \sqrt{V(\phi )} \left[1+O\left(\frac{1}{\phi } \right)\right],\,\, r\to r_s.
\end{equation}  

 Then we use \eqref{exp-beta}, \eqref{GrindEQ__8_} to get from \eqref{GrindEQ__6_}
\[\frac{d\phi }{dr} =-\frac{\sqrt{V(\phi )} }{4\pi r_s \Phi (\phi )} \simeq -\frac{n}{4\pi r_s } \phi ^{2n-1} .\] 

The solution is 
\[\phi (r)=\left\{\frac{n(n-1)}{2\pi } \frac{r-r_{1} }{r_s} +\frac{1}{\phi _{1}^{2(n-1)} } \right\}^{-\frac{1}{2(n-1)} } ,\quad \phi_1=\phi(r_1).          \] 

The limit (\ref{inf_1}) occurs if  \begin{equation}
\label{def_r0}
    \frac{n(n-1)}{2\pi }\cdot \frac{r_{1} -r_s }{r_s } =\frac{1}{\phi _{1}^{2(n-1)} } 
\end{equation}

Then
\begin{equation} \label{GrindEQ__9_} 
\phi (r)\sim \left\{\frac{2\pi r_s }{n(n-1)(r-r_s )} \right\}^{\frac{1}{2(n-1)} } .                                           
\end{equation} 

Using \eqref{exp-beta} we have
\begin{equation}\label{beta_ass}
\beta (r)\sim -2\ln \Phi (\phi )\sim -\left[\frac{2\pi r_s }{n(n-1)(r-r_s )} \right]^{n/(n-1)}.
\end{equation}

 The Kretschmann invariant near $r_s$ has the form
\begin{equation}
    R_{\alpha \beta \gamma \delta}R^{\alpha \beta \gamma \delta}\sim \frac{e^{-2\beta(r)}}{\left(r-r_s\right)^{\frac{2(2n-1)}{n-1}}}.
\end{equation}

The main outcome of these considerations is that there  exist singularities of solutions to systems (\ref{EE3})--(\ref{EE2}) for some non-zero value of the radial variable in curvature   coordinates, that is, SS are indeed possible. This does not mean that all the solutions have such singularities. It easy to see that for $n>1$ to have the singularity at some $r=r_s >0$,  the~condition (\ref{def_r0}) must be fulfilled (i.e., $\phi_{1}$ must be sufficiently large). The~latter depends on the initial conditions at infinity and this must  be derived numerically. This is a subject of the next~section.

\section{Numerical~Solutions}\label{numerical_solutions}
Here we restrict ourselves to the case of a long-range field $n>2$. Correspondingly,  taking into account (\ref{asympt_inf}), we assume the  conditions for the field 
\begin{equation}\label{asPhi}
 ~~\mathop{\lim }\limits_{r\to \infty } \left[r^{2} \frac{d\phi }{dr} \right]=-Q                 \end{equation} 
 yielding
\begin{equation} \label{asPhi} 
\mathop{\lim }\limits_{r\to \infty } \left[r\phi (r)\right]=Q.              
\end{equation} 

It was shown in~\cite{stashko2021accretion} that there is a unique solution of the problem for sufficiently large $r$ satisfying (\ref{EE3})--(\ref{EE2}) and (\ref{asPhi}). 
 The iteration procedure yielding the solution is described in~\cite{stashko2021accretion} for a monomial potential and it can be applied in case of (\ref{potential}) for sufficiently large $r$ (small $\phi$). This yields initial conditions for the ordinary differential system (\ref{EE3})--(\ref{EE2}) at some (large) $r_{init}$. Instead, one can directly use asymptotic expansions to derive the solution for large $r_{init}$.  We obtained the solution numerically by integrating the equations backward from $ r_ {init} $ to smaller values of $ r $, either  to a spherical  singularity or  to a point-like naked singularity at the origin. The~occurrence of singularity can be checked by means of  relation~\mbox{(\ref{def_r0})}; this can be used in order to define more precisely the singularity~radius. 
  
\begin{figure}[h]
    \includegraphics[width=85mm]{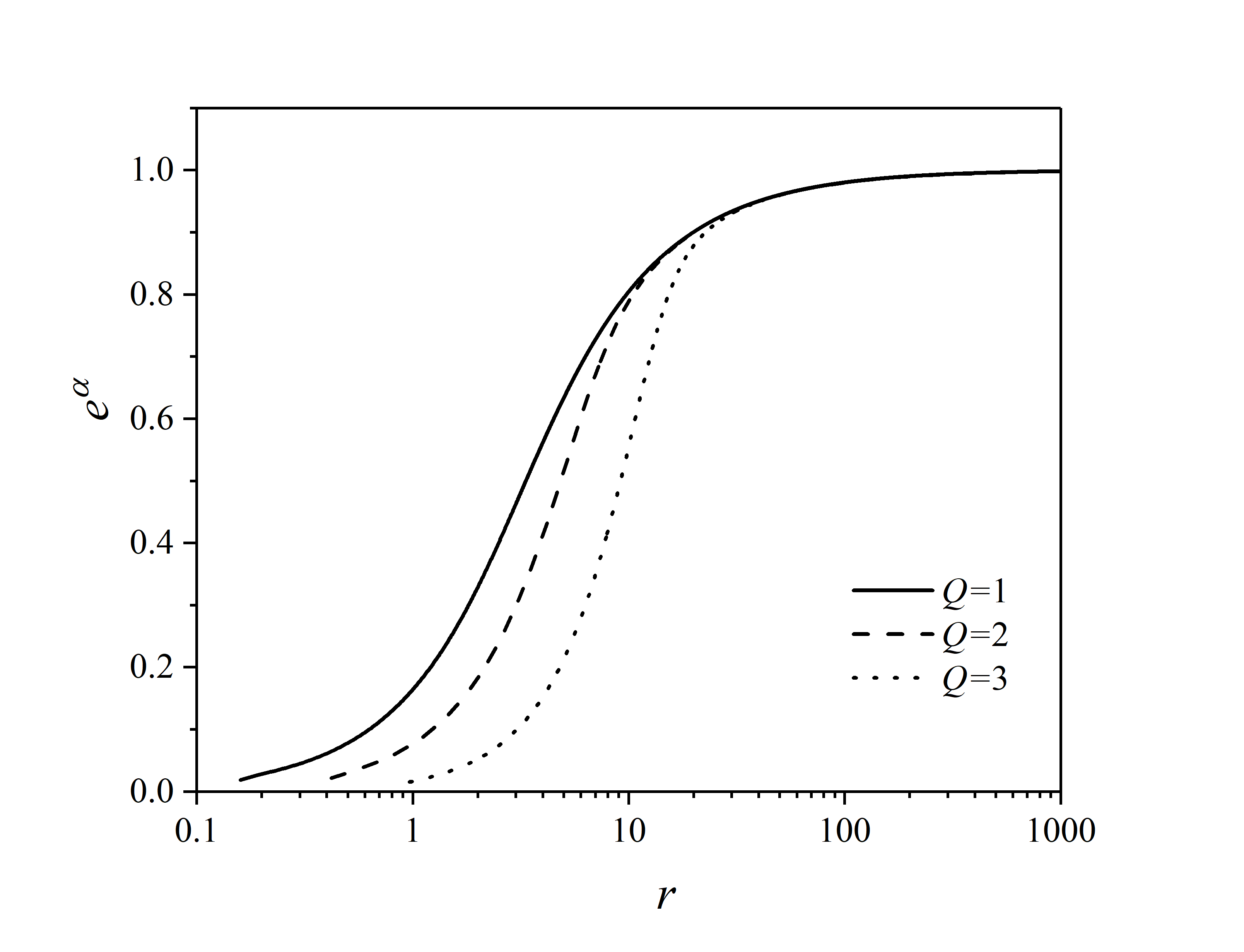}
    \includegraphics[width=85mm]{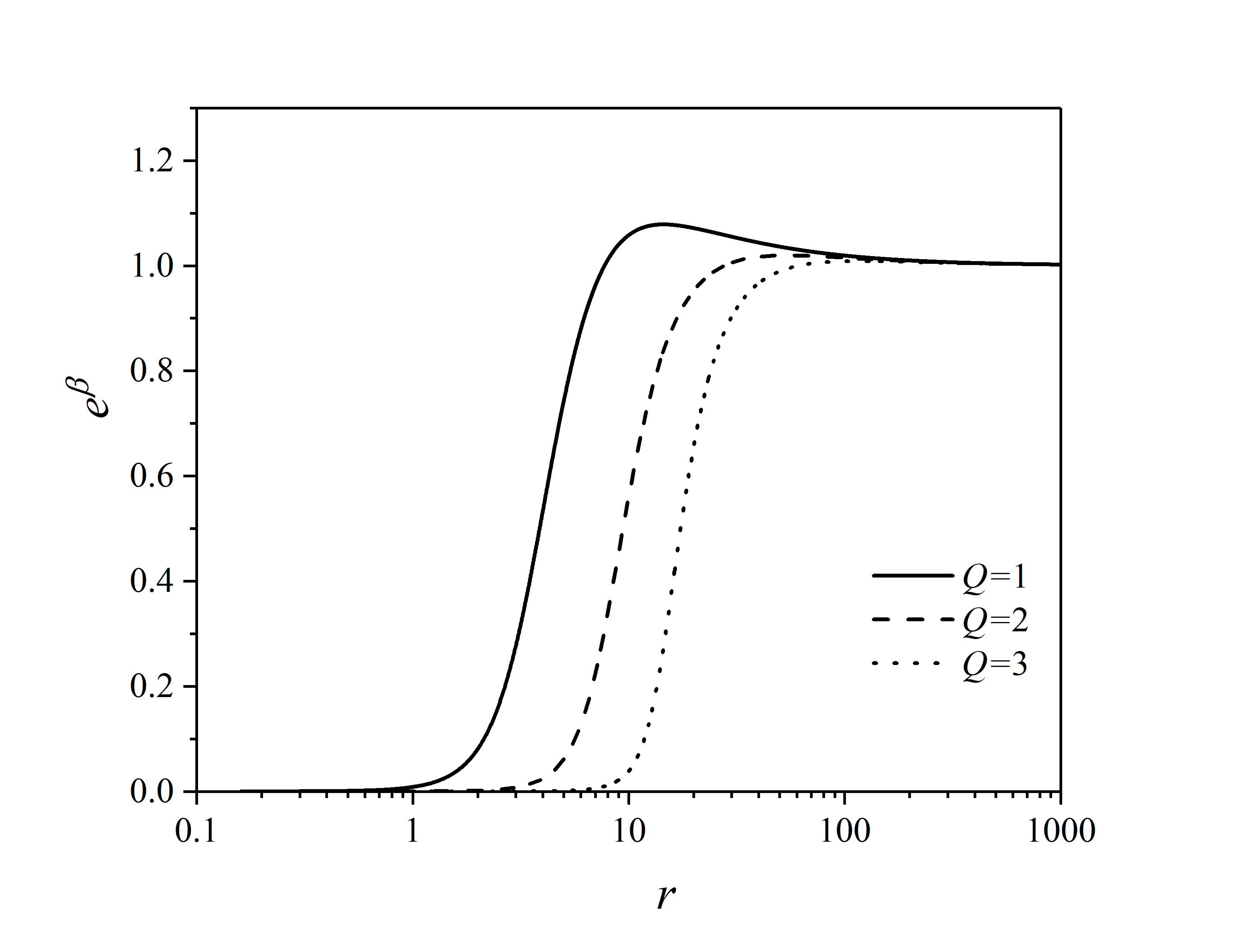}
    \caption{Behavior of metric functions $e^{\alpha}$ (\textbf{left}) and $e^{\beta}$ (\textbf{right}) for $M=1$ and different $Q$. }
    \label{fig:my_label0a}
\end{figure}

\begin{figure}[h]
    \includegraphics[width=85mm]{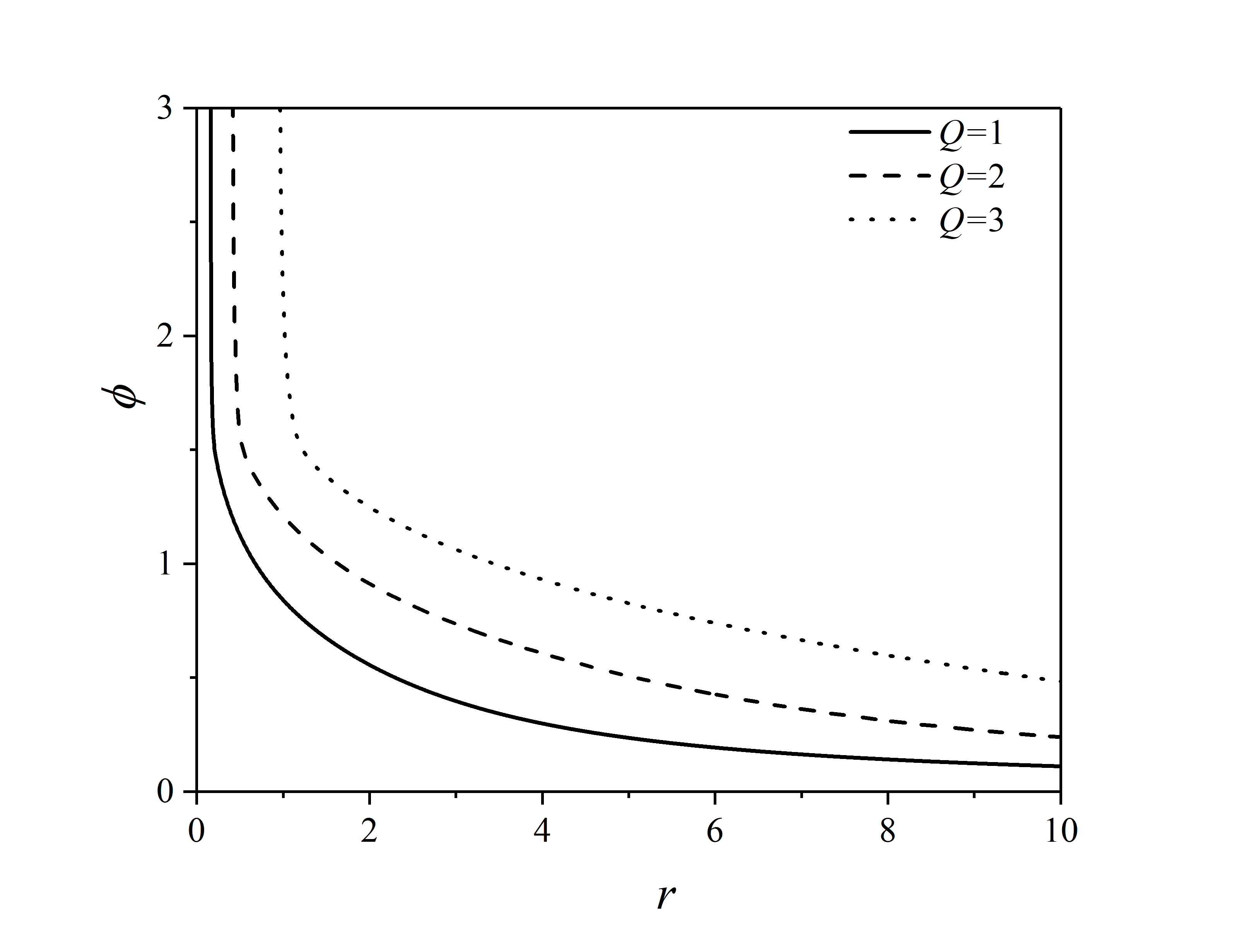}
    \includegraphics[width=85mm]{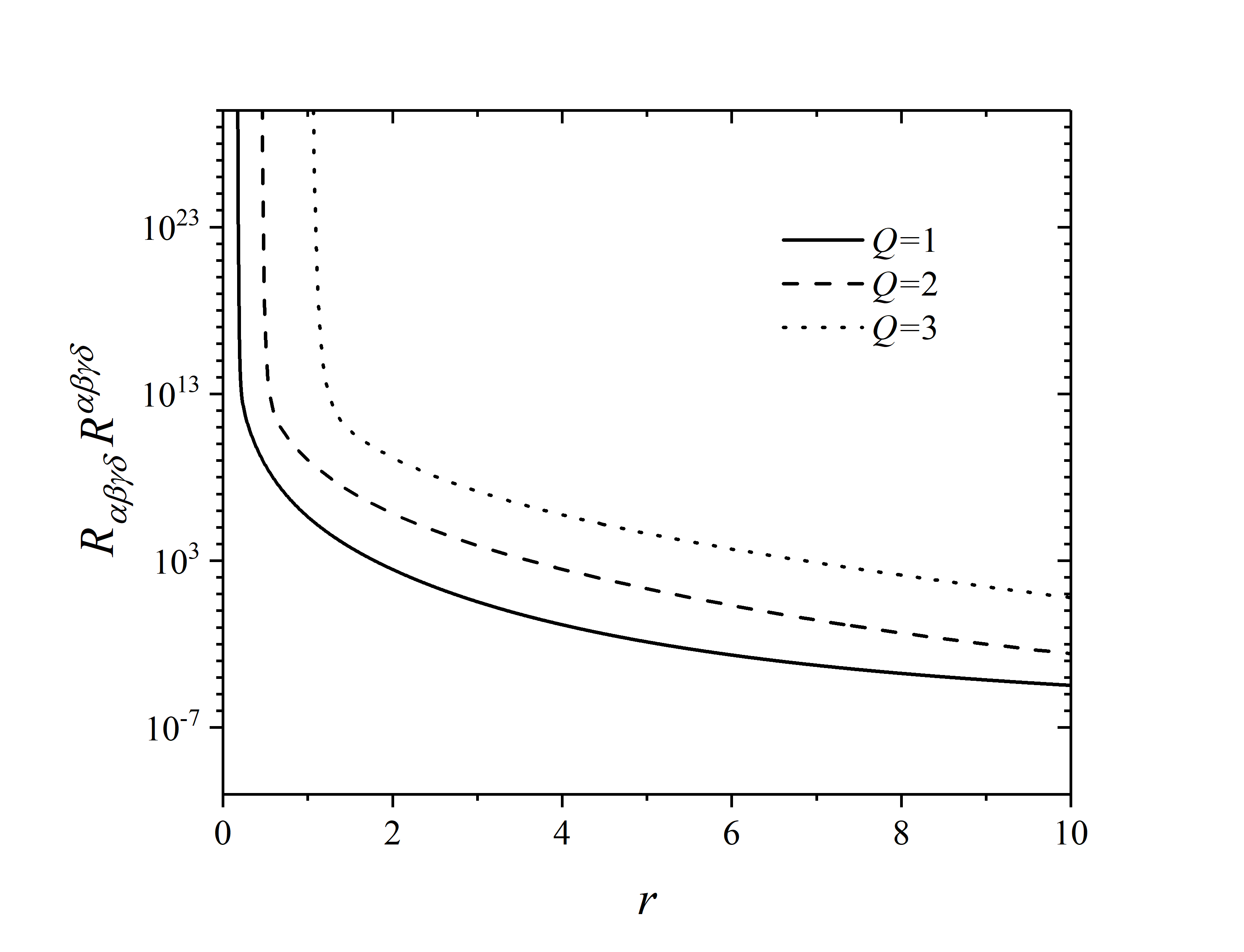}
    \caption{Behavior of SF 
 (\textbf{left}) and Kretschmann  invariant  $R_{\alpha\beta\gamma \delta}R^{\alpha\beta\gamma \delta}$ (\textbf{right})  for the same \mbox{parameters}.}
    \label{fig:my_label0b}
\end{figure}

\begin{figure}[h]
  
    \includegraphics[width=85mm]{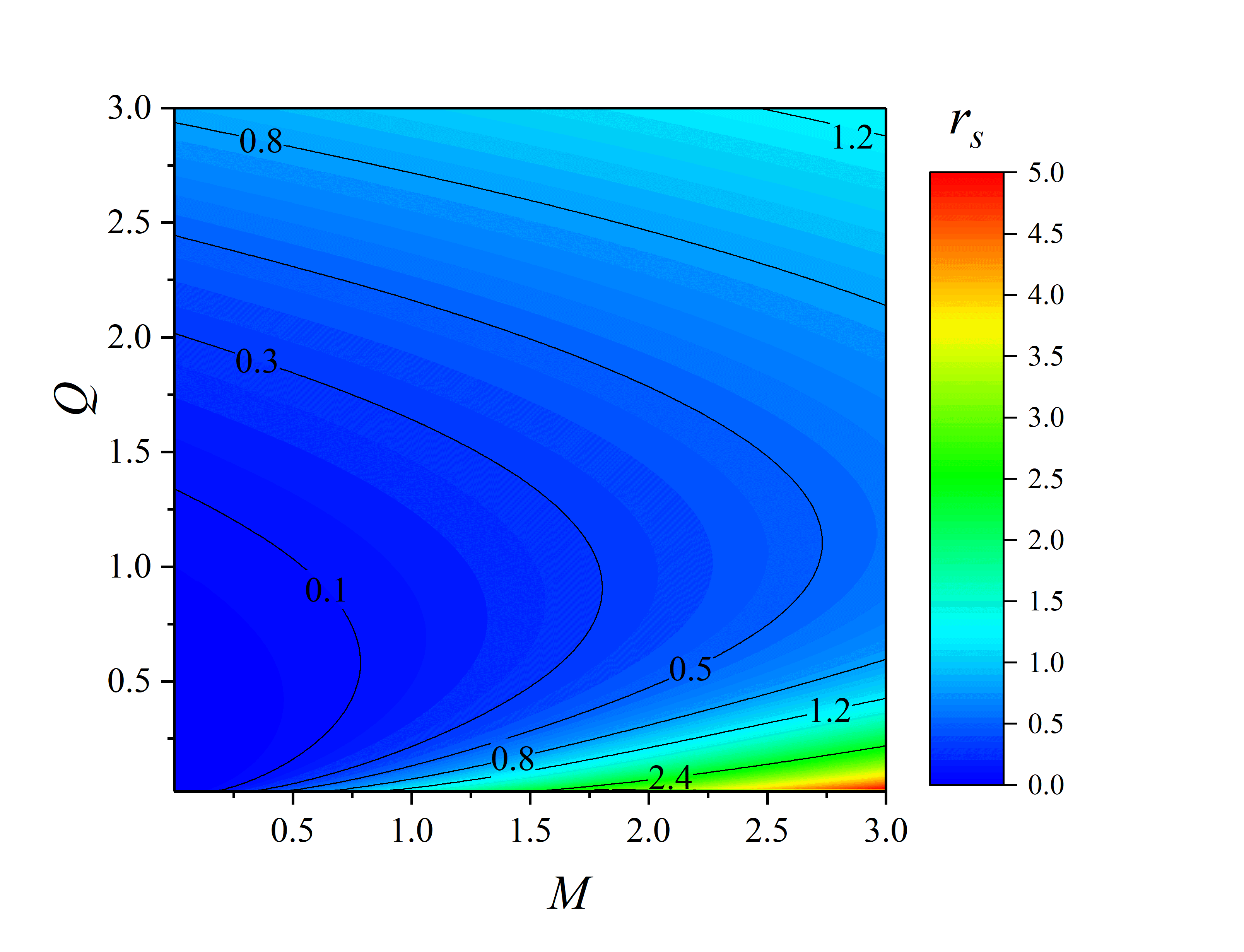}
    \includegraphics[width=85mm]{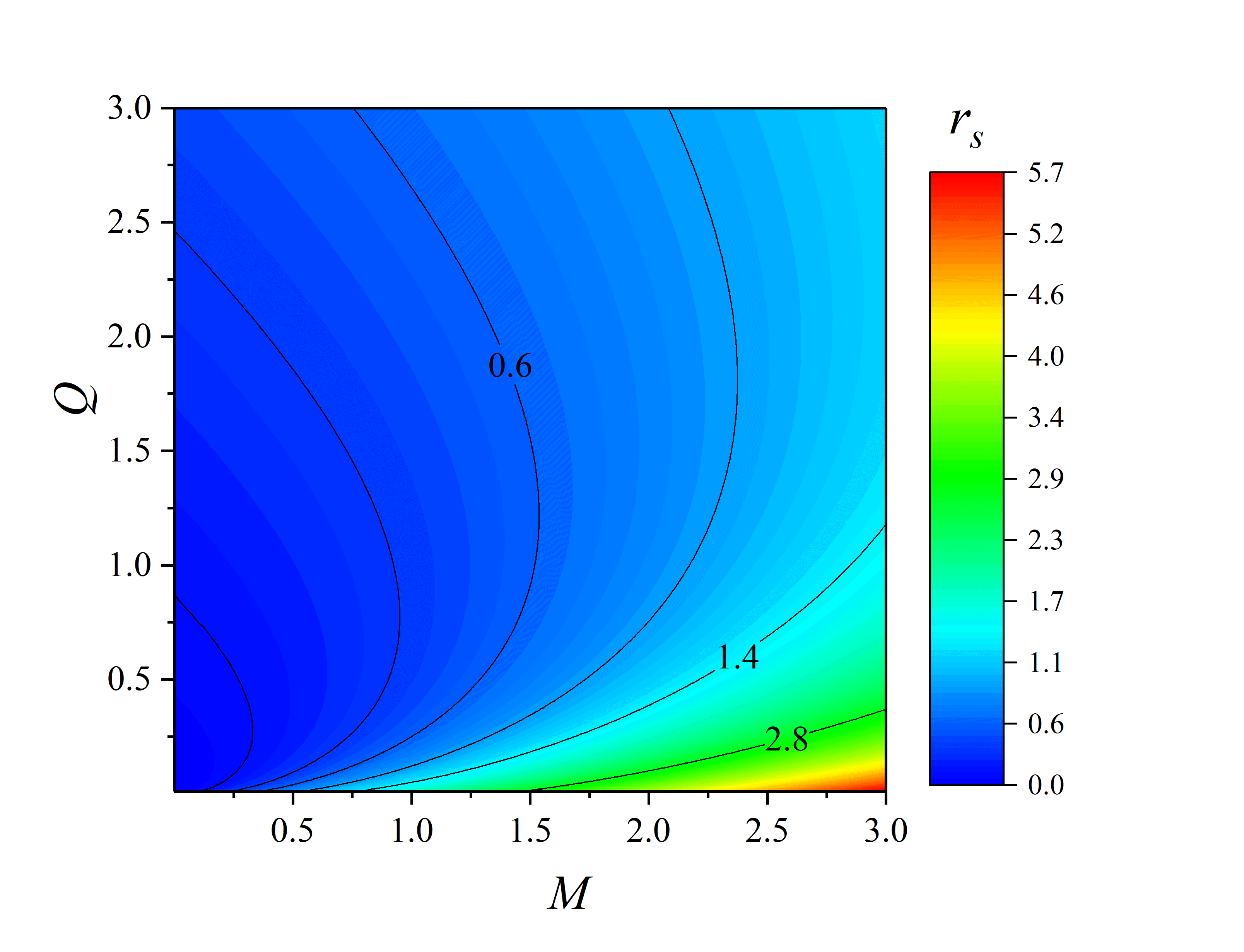}
    \caption{Radii of SS for different $M,Q$; $n=6$ (\textbf{left}), $n=18$ (\textbf{right}).}
    \label{fig:my_label1}
\end{figure}

\begin{figure}[h]

    \includegraphics[width=85mm]{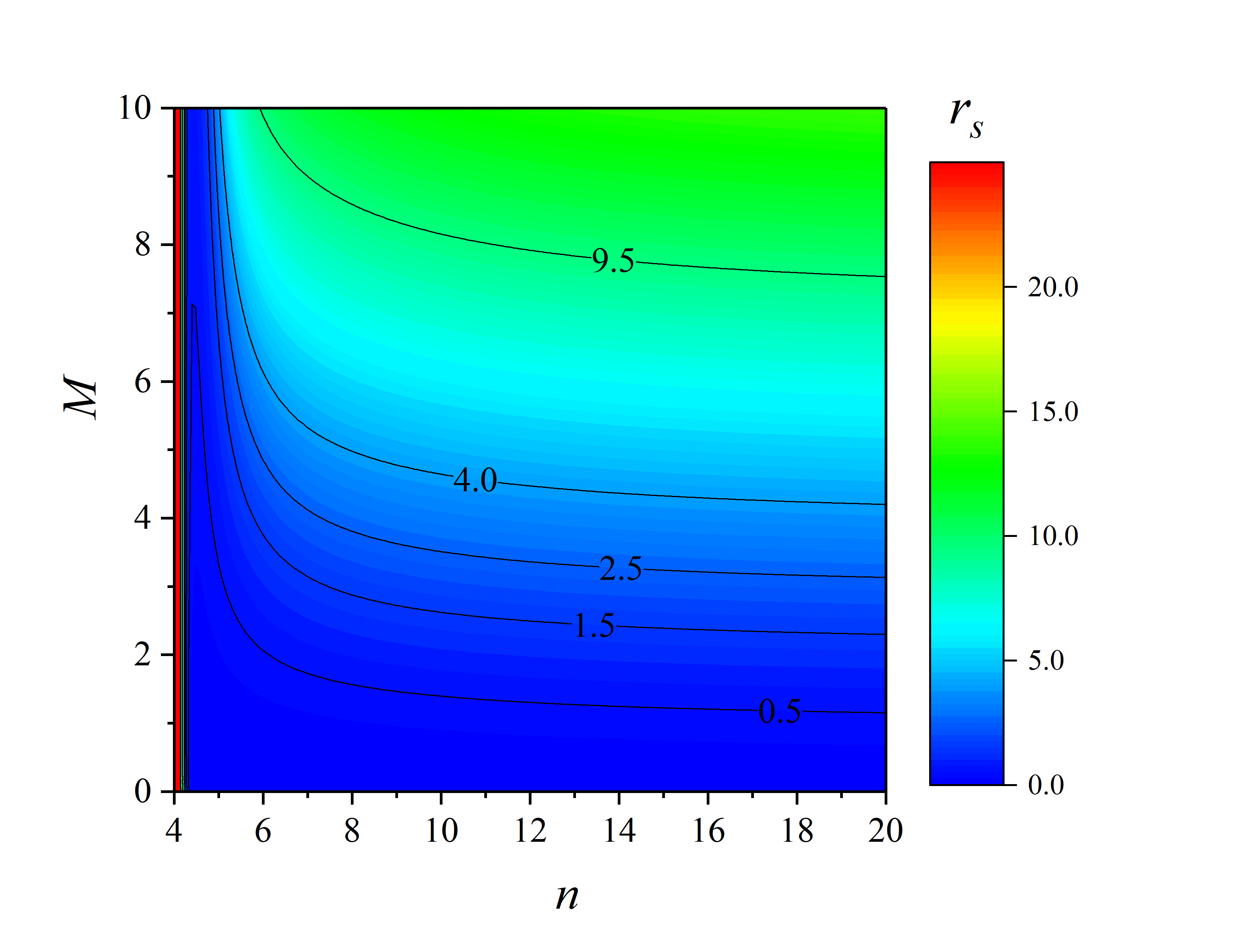}
    \includegraphics[width=85mm]{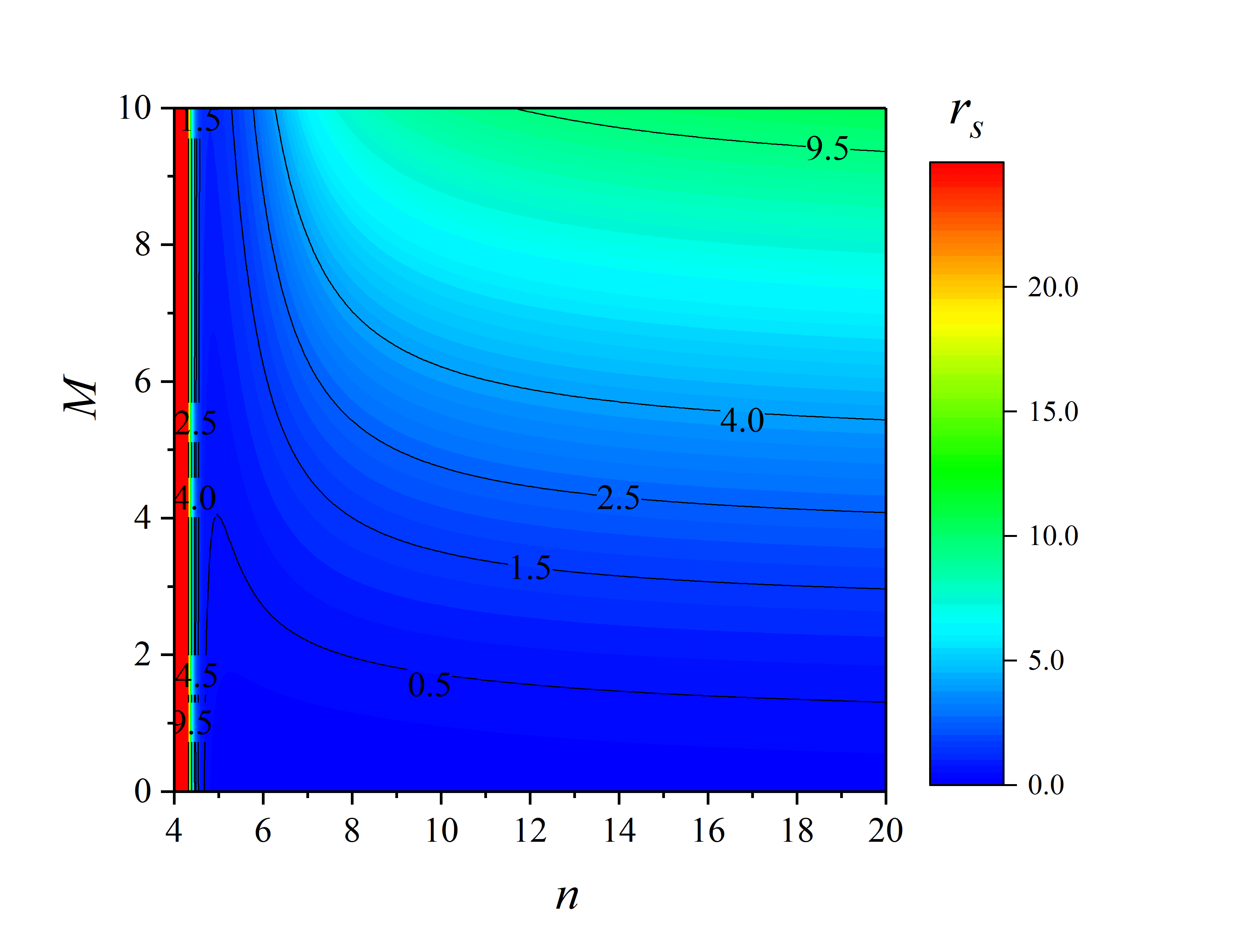}
    \caption{Radii of SS for different $n,M$; $Q=0.5$ (\textbf{left}), $Q=1$ (\textbf{right}).}
    \label{fig:my_label2}
\end{figure}

\begin{figure}[h]
  
    \includegraphics[width=85mm]{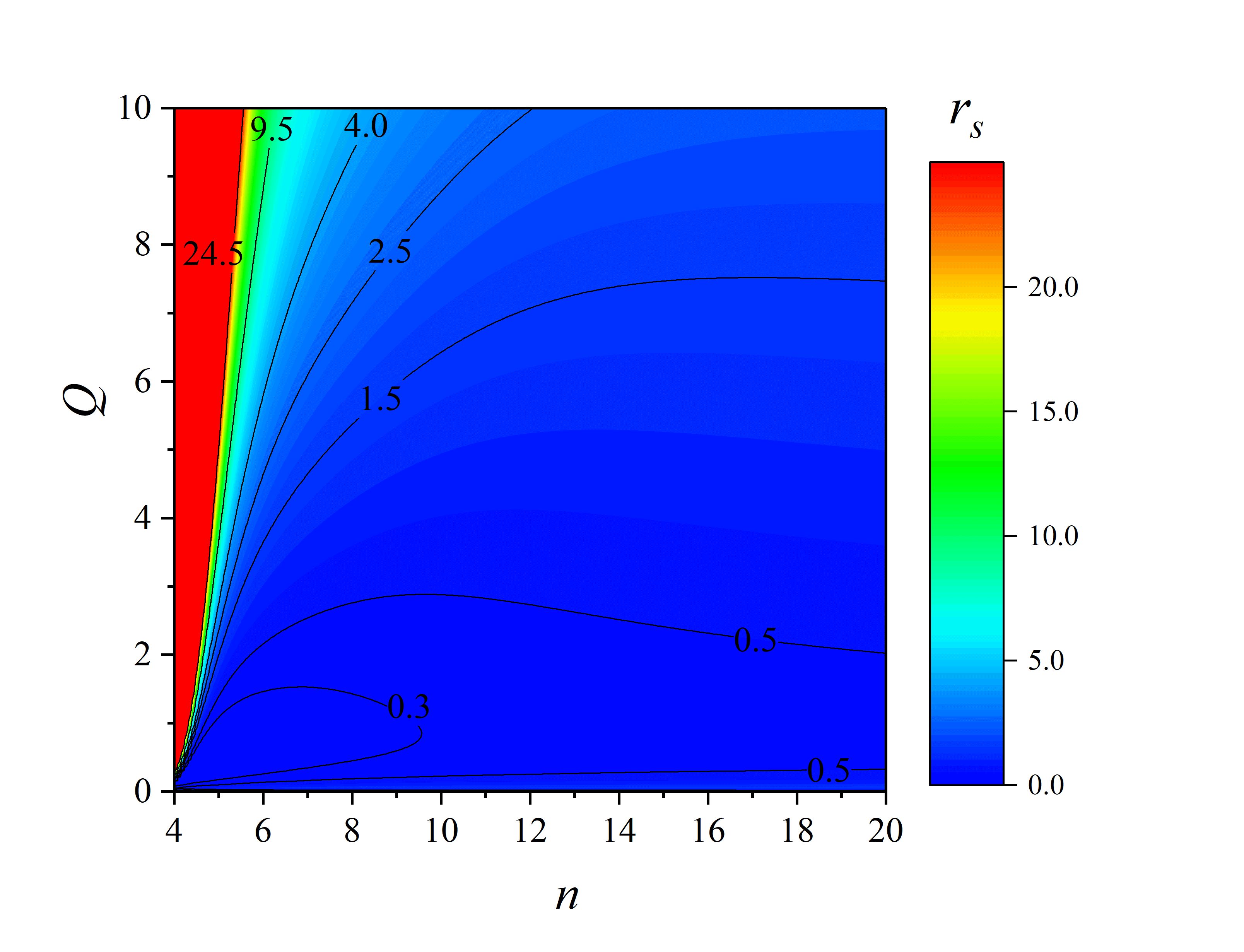}
    \includegraphics[width=85mm]{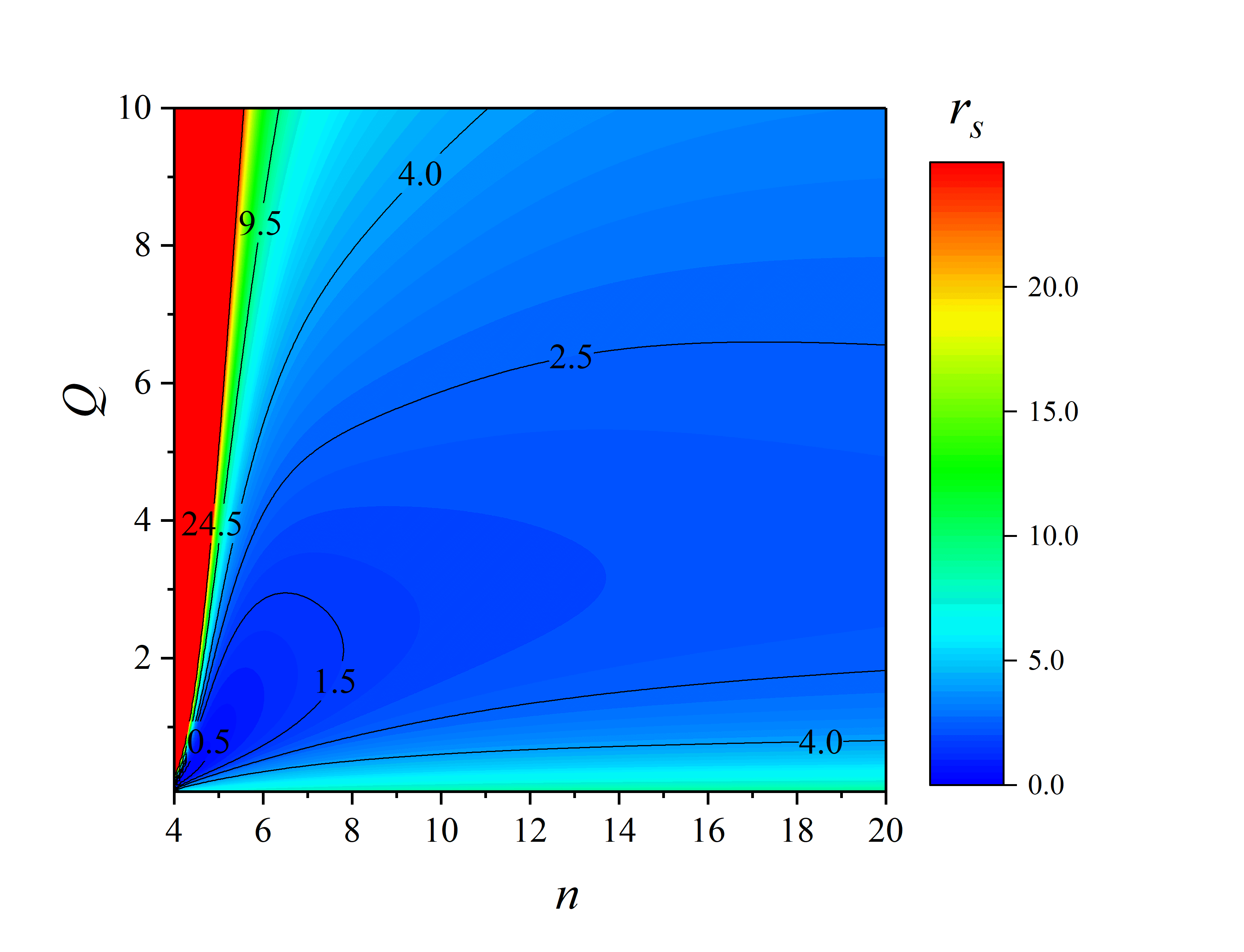}
    \caption{  Radii of SS for different $n,Q$; $M=1$ (\textbf{left}) and $M=5$ (\textbf{right}).}
    \label{fig:my_label3}
\end{figure}

  Figures~\ref{fig:my_label0a} and \ref{fig:my_label0b} show typical behavior of solutions in the case of SS; we see that 
 $e^{\alpha}$ is    monotonically increasing,  whereas $e^{\beta}$ reaches a maximum and then decreases to one for $r\to\infty$. Additionally, we note that near the singularity $e^{\beta-\alpha}\to 0$ for $r\to r_s+0$ in accordance with Section~\ref{asymptotic_behavior}. Figures~\ref{fig:my_label1}--\ref{fig:my_label3} show a non-trivial dependence of the singularity radii $r_s$ upon parameters $n,M,Q$. 

\unskip
\vspace{-9pt}

\unskip
\unskip
\unskip
\section{Test Particle~Motion}\label{Test_particle_motion}
In this section, we consider the massive test particles motion around static spherically symmetric configurations with SF to describe in order to describe  characteristics of accretion onto a singularity. We use the Page--Thorne model~\cite{Page_Thorne} of a thin accretion disk  (AD). In~this model, the~averaged motion of the accreting matter is essentially described by SCOD in the equatorial~plane.

In the case of spherically symmetric space-time with metric (\ref{metric}), the~  standard procedure  yields the first integrals for test particle trajectories in the equatorial plane $\theta=\pi/2$ ($\tau$ is a canonical parameter.): 
\begin{equation}
 \label{geodesics_1}
 e^{\alpha}\left(\frac{dt}{d\tau}\right)^2 - e^{\beta}\left(\frac{dr}{d\tau}\right)^2  - r^2\left(\frac{d\phi}{d\tau}\right)^2=S\,,
 \end{equation}
\begin{equation}
 \label{geodesics_2}
 e^{\alpha}\left(\frac{dt}{d\tau}\right)=E, \quad  r^2\left(\frac{d\phi}{d\tau}\right)=L\,,
 \end{equation}
 where $S=0$ in case of null trajectories and $S=1$ for the test particles with the non-zero mass; $L,\,E$ are the integrals  of motion.
 This yields
\begin{equation}
 \label{1D particle}
 e^{\alpha+\beta}\left(\frac{dr}{d\tau}\right)^2=E^2-
 U_{\rm eff}(r,L,S)  \,,
 \end{equation}
 where effective potential $U_{\rm eff}(r,L,S)=e^{\alpha}\left(S+{L^2}/{r^2}\right)$.
 
\begin{figure}[h]
 
    \includegraphics[width=110mm]{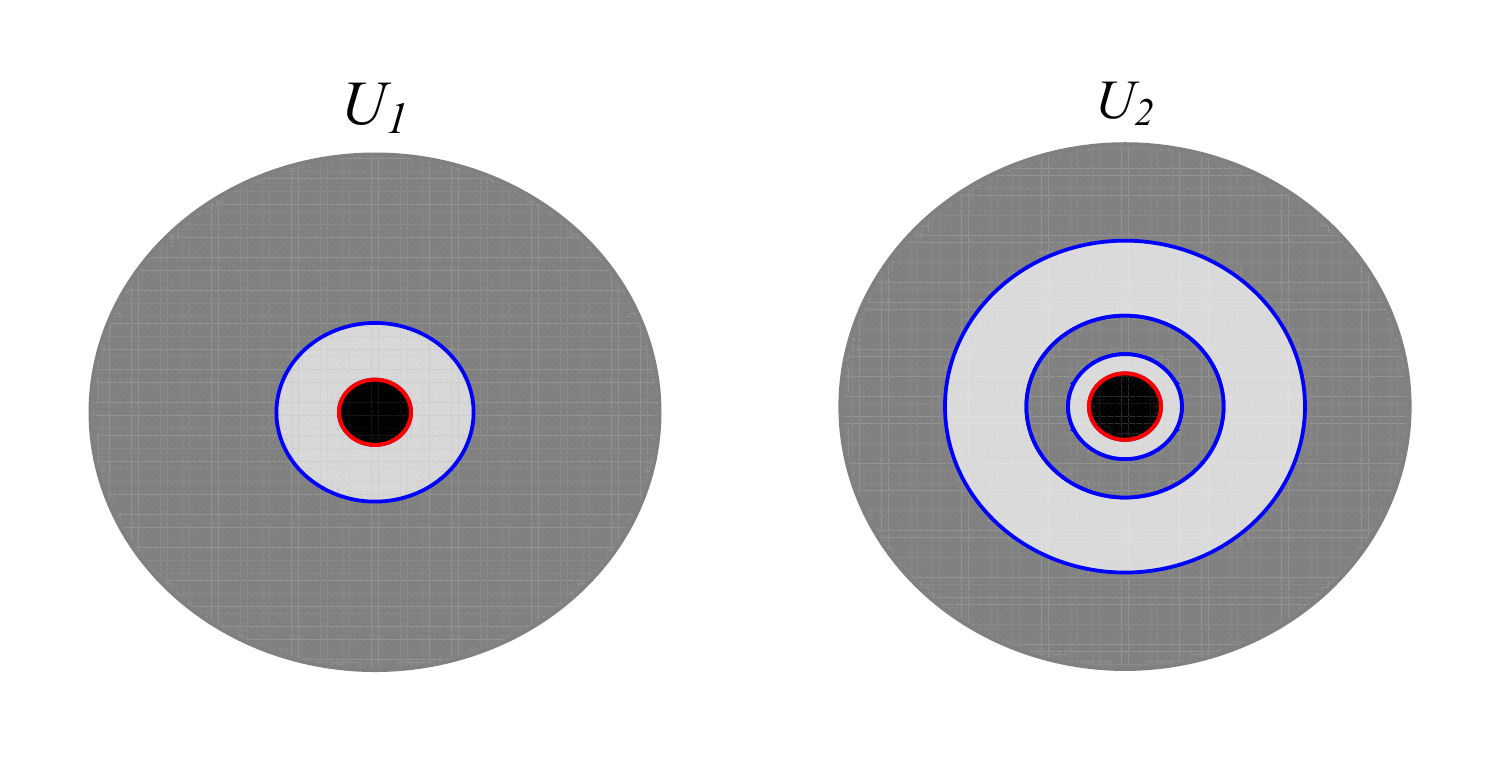}
    \caption{The schematic examples of possible SCODs in the equatorial plane. In~both cases, black spots in the center represent SS at finite values of $r$. Dark and light gray rings correspond to the sequences of SCO and unstable circular orbits correspondingly. Blue circles show the boundary radii of SCOD as described in Table \ref{tab1}.}
    \label{fig:example}
\end{figure}

 The form of $U_{\rm eff}(r)$, in~particular, the disposition of its  minima and maxima, defines the distribution of stable circular orbits (SCO) and unstable ones. The~SCO  distribution (SCOD)  is the most important because it forms  the basis for evaluating the properties of an accretion disc around the configuration described by Equations~(\ref{EE3})--(\ref{EE2}). Possible types of SCODs are listed in Table \ref{tab1} and described schematically on Figure \ref{fig:example}.
 \begin{table}[h]
    \centering
    \begin{tabular}{|c|c|c|c|}
\hline
\textbf{Type}	& \textbf{$r_{\rm stable}$}                &  $r_{\rm unstable}$                & Photon sphere\\
\hline
$U_1^{(-)}$		& $(r_1,\infty)$      	    	&   $(r_s,r_1)$                 &   $-$\\
\hline
$U_1^{(+)}$		& $(r_1,\infty)$                &  $(r_s,r_1)$                  &   $+$\\
\hline
$U_2$   		& $(r_1,r_2)\cup(r_3,\infty)$     &   $(r_s,r_1)\cup(r_2,r_3)$    &$-$\\
\hline
\end{tabular}
\caption{Possible types of SCOD\label{tab1}}
\end{table}

\begin{figure}[h]

    \includegraphics[width=90mm]{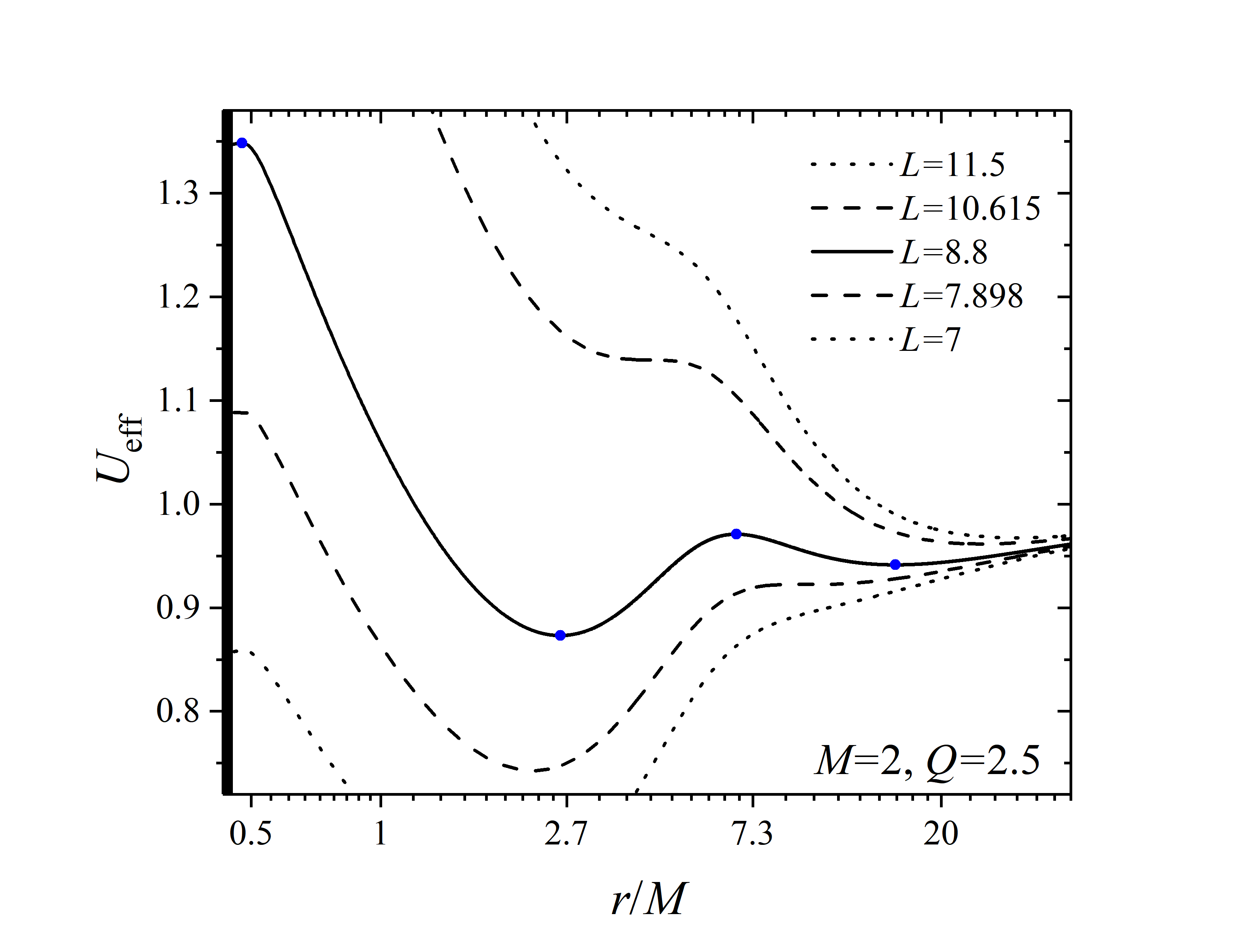}
    \caption{Typical examples of $U_{\rm eff}$ for configuration with $U_2$  SCOD type. The~blue points show the corresponding extrema. There is only one minimum for large $L$.  }
    \label{fig:7}
\end{figure}

We use a method of our works~\cite{2018Stashko,Stashko_Zhdanov_2019a} to study bifurcations associated with the appearance and disappearance of the minima of $U_{\rm eff}$. Figure \ref{fig:7} illustrates how the shape of the effective potential changes with an increase in $L$ in case  of the $U_2$  SCOD type. Essentially this is connected with investigation of joint  conditions $U'_{\rm eff}=0$ and $U''_{\rm eff}=0$, which allow us to exclude $L$; this leads to a necessary condition $F(r)=0$, where
\begin{equation} \label{bifurcation_condition}
  F(r)= r \alpha''(r)-r \alpha'(r)^2+3 \alpha'(r).
\end{equation}

For the congruence of circular orbits with different radii in  equatorial plane  we get  dependencies of the specific energy and  the specific angular momentum, and~the angular velocity   $\Omega=d\varphi/dt$ upon radius $r$ as follows
\begin{equation}
\label{tilde_E_L_Omega(r)}
  \tilde E^2(r)=\frac{2e^{\alpha(r)}}{2-r\alpha'(r)},~~ \tilde L^2(r)=\frac{r^3\alpha'(r)}{2-r\alpha'(r)},~~\Omega^2(r)=\frac{\alpha'(r)e^{\alpha(r)}}{2r}
\end{equation}

\begin{figure}[h]
   
    \includegraphics[width=95mm]{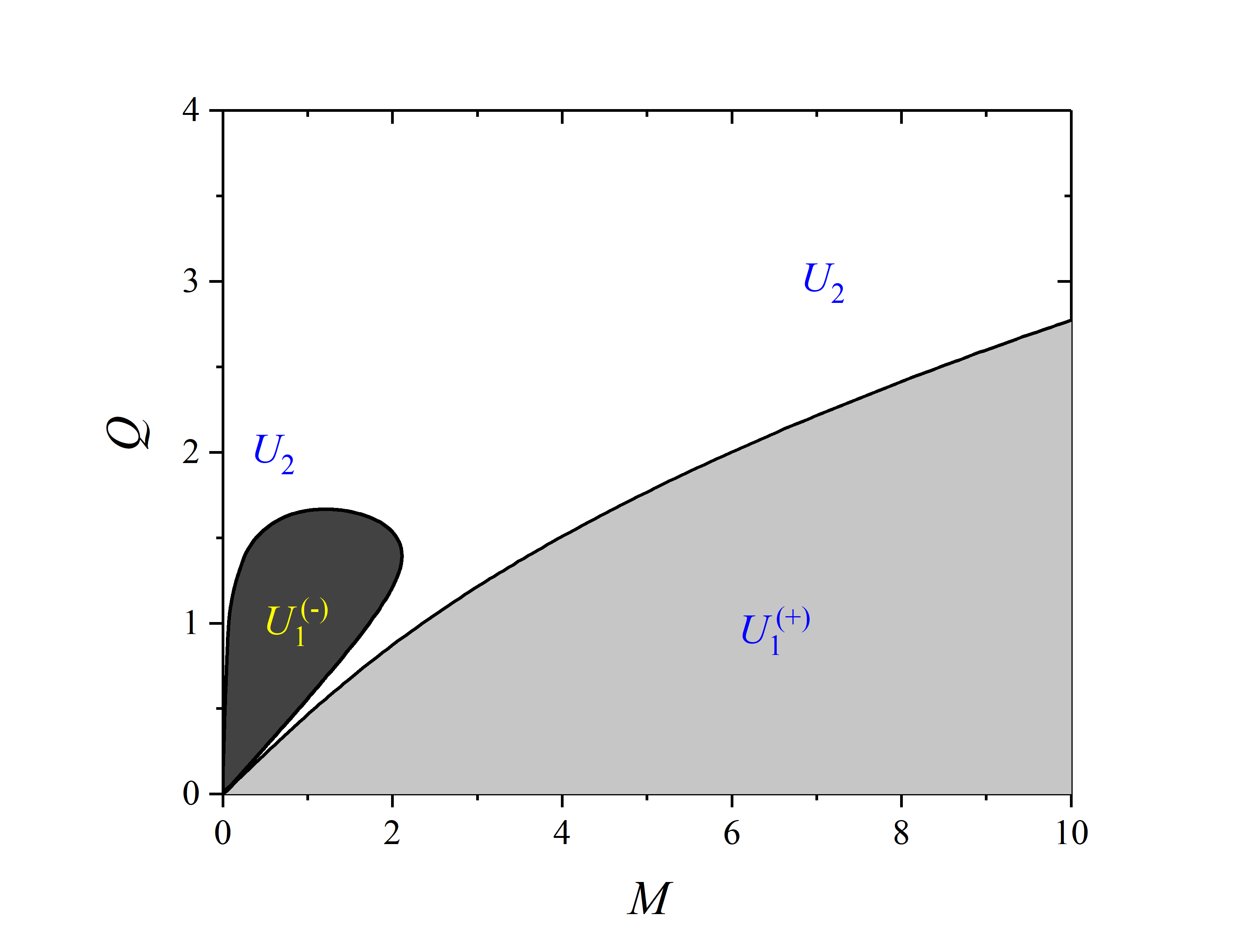}
    \caption{Domains of parameters on $(M,Q)$ plane for $n=3$. Every point of this plane corresponds to a solution of Equations~(\ref{EE3})--(\ref{EE2}) with conditions (\ref{asy}), (\ref{asympt_inf}). The~SCOD type was determined for every such solution; gray, dark gray and white colors correspond to the $U_1^{(+)}$, $U_1^{(-)}$ and $U_2$ type, respectively.}
    \label{fig:8}
\end{figure}

\begin{figure}[h]
  
    \includegraphics[width=85mm]{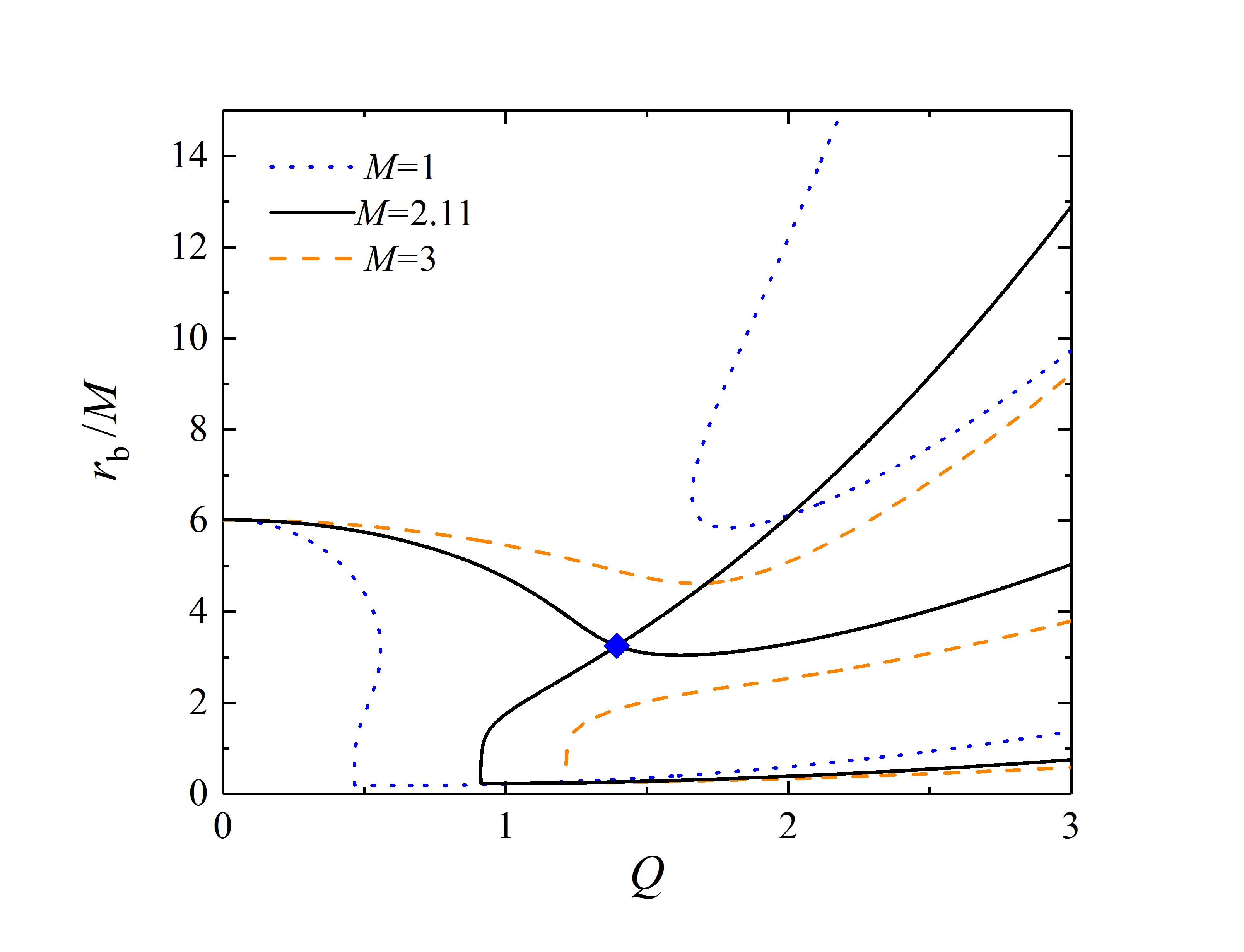}
    \includegraphics[width=85mm]{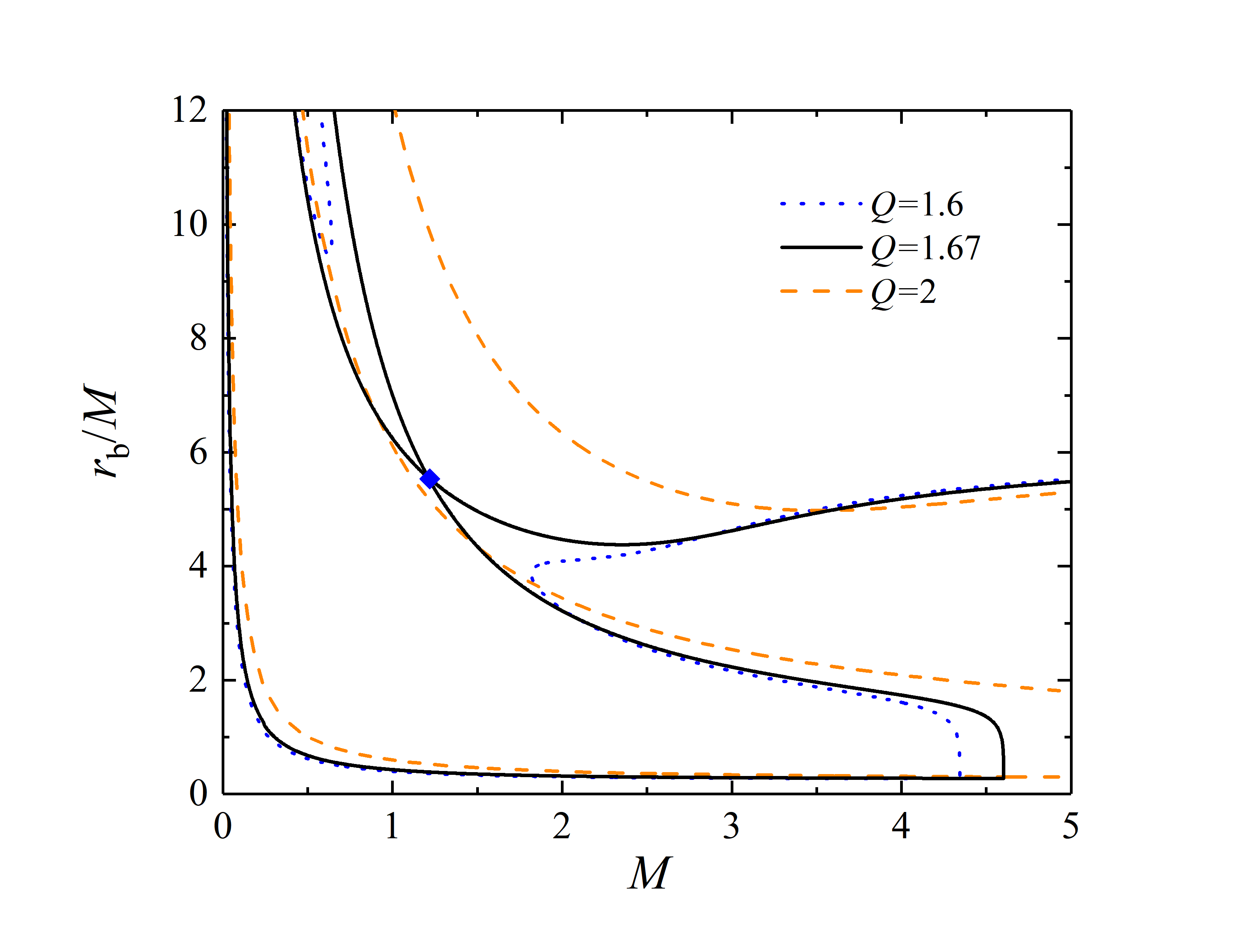}
    \caption{Boundary radii $r_b/M$ of SCO regions as  functions of $Q$ (\textbf{left}) and $M$ (\textbf{right}) for several values of $M$ and $Q$. The~corresponding points where curves qualitatively change their behavior by reconnection are  $(M,Q)\simeq(2.11,1.39)$ and $(M,Q)\simeq (1.2,1.67)$ for left and right pictures, respectively and shown by blue~squares.}
    \label{fig:9}
\end{figure}

Using Equation~(\ref{bifurcation_condition}), we numerically get the bifurcation values $r_b, \, L^2_b=\tilde L^2(r_b)$ and $E^2_b=\tilde E^2(r_b)$ under conditions  that  $\tilde E^2(r_b)>0$ and $\tilde L^2(r_b)>0$. This allows us to determine SCOD types  for given configuration parameters. Figure \ref{fig:8} shows the corresponding results in the  $M-Q$ plane. Also, Figure \ref{fig:9} illustrates the dependence of the boundary radii (i.e., radii of the blue circles in Figure \ref{fig:example}) upon $M$ and  $Q$. 
\unskip
\unskip

\begin{figure}[h]
     \includegraphics[width=85mm]{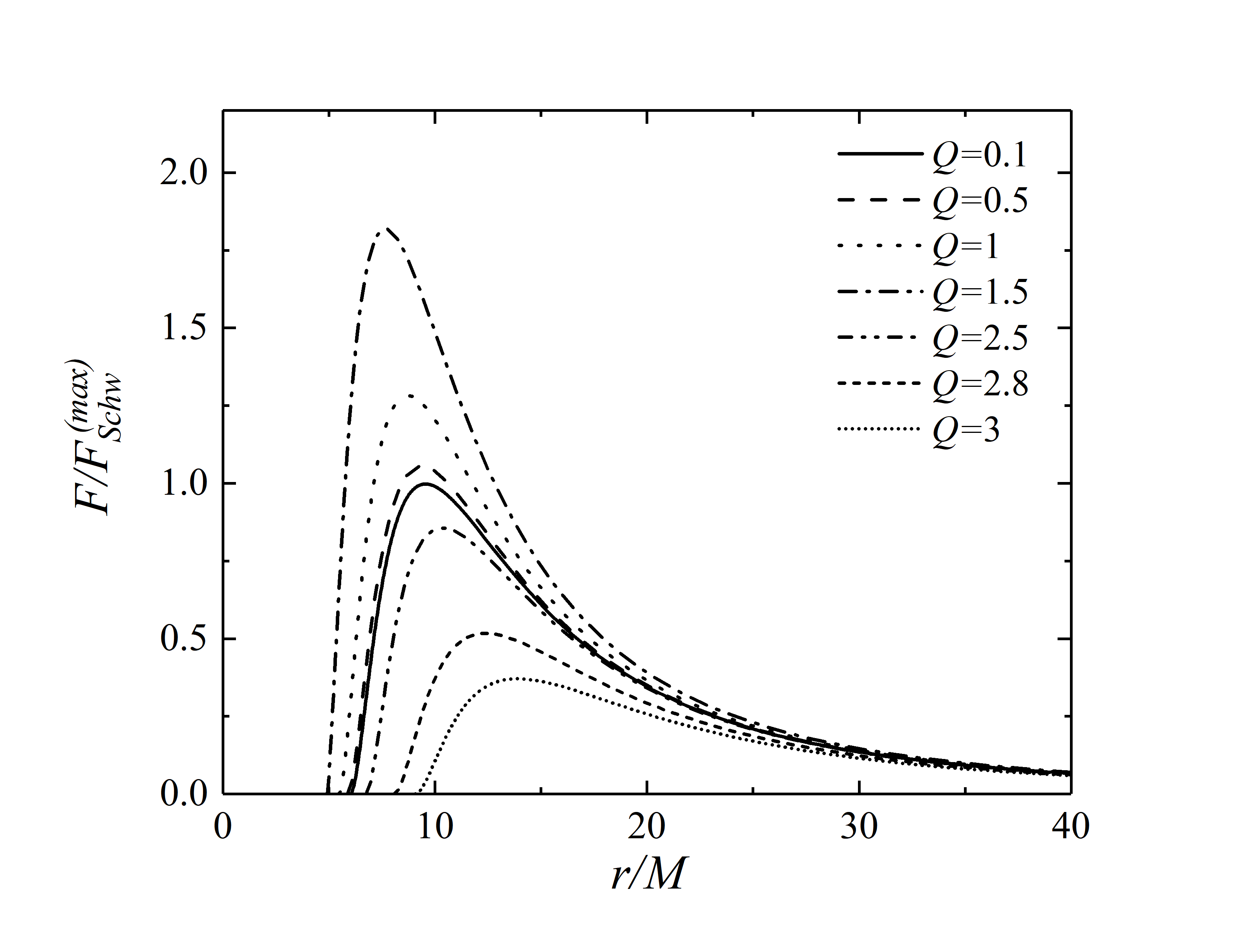}
    \includegraphics[width=85mm]{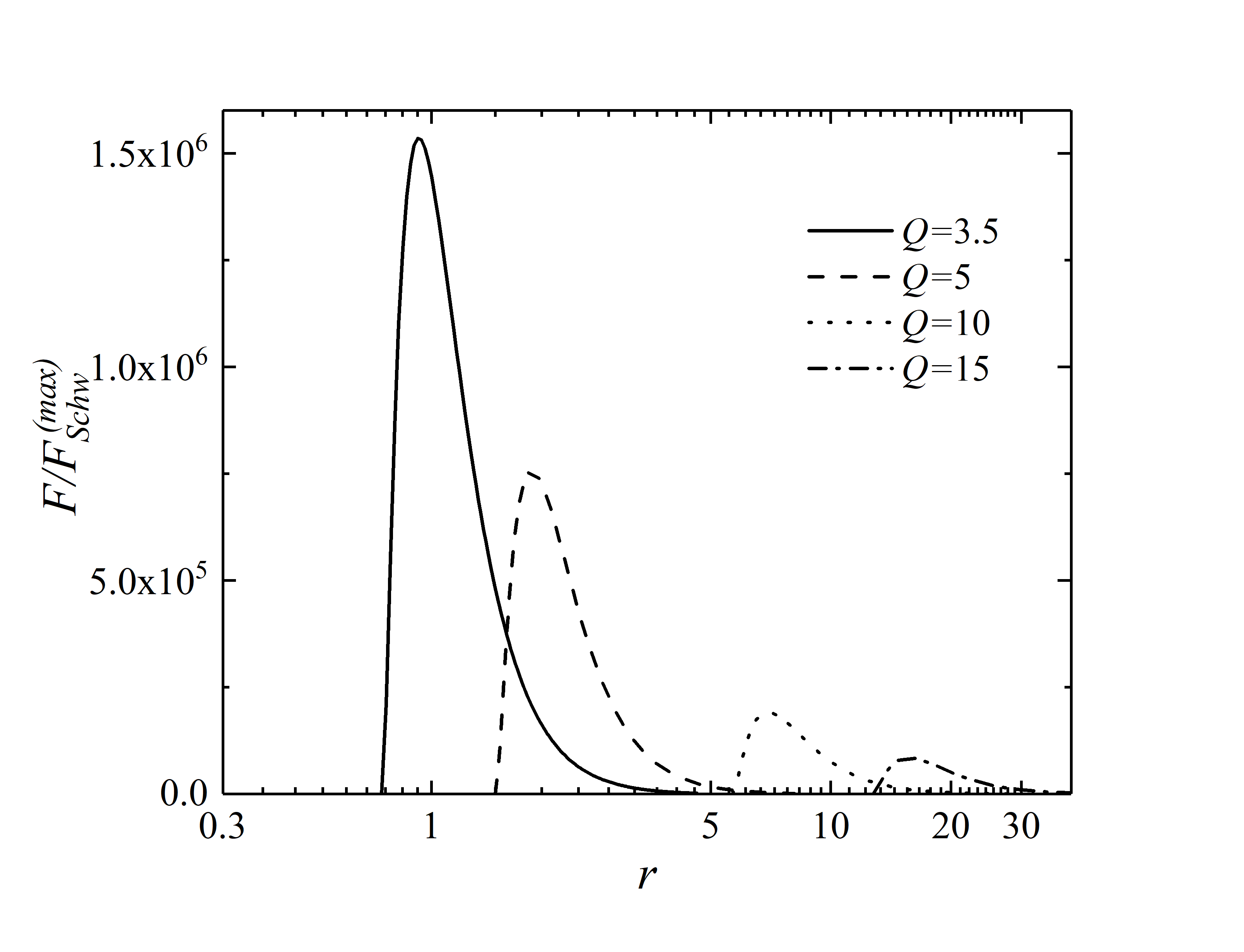}
    \caption{The  normalized flux ($M=4$) from inner (\textbf{right}) and outer (\textbf{left}) parts of~AD.}
    \label{fig:Flux}
\end{figure}

\unskip
Now we consider radiation from the stationary thin AD described by Page--Thorne model~\cite{Page_Thorne}. The~ time averaged radiation flux $F(r)$  from the surface of the accretion with inner edge located at the boundary SCO  $r=r_{\rm b}$ is
\begin{equation}
F(r)=-\frac{\dot{M_0}}{4\pi\sqrt{|{}^{(3)}g|}}\frac{\Omega_{,r}}{\left(E-\Omega L\right)^2}\int\limits_{r_{\rm b}}^{r}\left(E-\Omega L\right)L_{,x}\,dx,
\end{equation}
where $\sqrt{|{}^{(3)}g|}=re^{(\alpha+\beta)/2}$ is the metric's determinant in the equatorial plane   and $\dot{M_0}$ is the mass accretion rate that is assumed to be constant. Specific particle energy, momentum, and angular velocity are defined from (\ref{tilde_E_L_Omega(r)}). 
We obtained the radiation flux for several values $Q$ in presence/absence of inner stable ring of AD. The~corresponding fluxes are shown in Figure~\ref{fig:Flux}. To~compare different types, we normalize them to the maximum flux in the case of a Schwarzschild black hole  $F_{\rm Schw}^{\rm (max)}\simeq 0.0001719\dot{M_0}/ 4\pi M^2$. We see that the maximum value of different types  can be less or more  then $1$ as distinct from the case of  the  massless scalar field~\cite{Chowdhury_2012,Gyulchev_2020},  where $F/F_{\rm Schw}^{\rm (max)}\geq1$ due to $r_b\leq 6M$. 
\unskip
\section{Discussion}\label{Discussion}
We have shown that General Relativity allows  the existence of SS  in static spherically symmetric configurations in the case of the SF potential  $V(\phi)= \sinh(\phi^{2n})$, which represents potentials with fast growth rate.  This statement  follows from the analytical reasoning of Section~\ref{asymptotic_behavior} and is confirmed by numerical simulations of Section~\ref{numerical_solutions}. These SS are ``physical'' singularities that cannot be removed by a coordinate transformation:  this can be seen from the behavior of  the 
Kretschmann invariant. These  are naked singularities, because~they can be observed by a distant observer; it is easy to verify this by considering the radial motion of photons, taking into account that near the singularity   $\exp(\beta-\alpha)\to 0$ for $r\to r_s+0$.


We note that the appearance of singularities in solutions of nonlinear equations is a fairly typical case. However, 
the situation with SS is different from the case of monomial or the other exponentially bounded potentials, when SS are suppressed by the gravitational field and we only have a naked singularity at the center~\cite{ZhdSt}.  In~the case of \mbox{$V(\phi)=\sinh(\phi^{2n})$}, the~sharp growth of the field near the singularity overcomes  this suppression. This suggests that SS can occur  for   more general potentials with fairly rapid~growth.  

Having the numerical solutions, we investigated the location of stable and unstable circular orbits. There are two main types of SCOD in our case: (i)  $U_1$ is  similar to SCOD in the case of the Schwarzschild metric, when SCO radii must be larger than some boundary value; (ii) $U_2$ type is formed by a non-connected distribution with two regions of SCO separated by a ring of unstable circular orbits. These distributions of the stable orbits are directly related to the structure of the thin accretion disk in the Page--Thorne model~\cite{Page_Thorne}. Perhaps this result can also be important for more complex AD models, if, of~course, the~above scalar fields do exist.   Note that these types of SCODs also arise in the case of solutions with monomial potentials~\cite{stashko2021accretion}. Additionally, it should be noted that $U_2$ type is not observed  for M87* shadow~\cite{Akiyama2019}. However, for~definite answer  observations with better resolution are~mandatory.

We have no answer to the question of whether strongly nonlinear fields really exist. Moreover, the~question remains, how the spherical singularity can form.  This is a question of the same order as  the origin of point  naked singularities, as~well as  other exotic structures such as bosonic stars, wormholes, etc.~\cite{2014CQGra..31s5013S,Shao_2021,Chakraborty_2019,2020PhRvD.101d3005B,Pugliese_2011,Pugliese_2013,Joshi_2013,Shahidi_2020,2016PhRvD..93b4024B,2018Stashko, Karimov_2019,paul2019observational,narzilloev2021particle,Abdujabbarov_2009,2014PhRvD..90b4071L, 2016Vincent,Grandclement_2014,LieblingL,Lamy_2018,Dymnikova_2019,Herdeiro_2021,Stuchl_k_2015}.     We note, however, that for some sets of configuration parameters, it could  be difficult to distinguish SS from  ordinary black holes relying upon the accretion disk structures. On~the other hand, for~other sets of parameters, we may have an unconnected SCO region, which would be different from the point of view of a distant observer.  Additional information on the  existence/non-existence of the strongly non-linear scalar fields  may come  from  considerations of early cosmological processes. So the hypothesis of the existence of the spherical singularities seems to be~testable.

\acknowledgments{This work is supported by National Research Foundation of Ukraine (project No. 2020.02/0073).}

\FloatBarrier
\bibliography{references.bib}
\end{document}